\documentstyle[twocolumn,prl,aps,epsfig]{revtex}
\title{Exact Multifractal Exponents 
for Two-Dimensional Percolation}
\begin{document}
\input epsf
\input mssymb
\draft
\author{Bertrand Duplantier}
\address{Service de Physique Th\'{e}orique de Saclay, F-91191 Gif-sur-Yvette 
Cedex,\\
and Institut Henri Poincar\'{e}, 11 rue Pierre et Marie Curie, 75231 Paris Cedex
05, France}
\date{November 23, 1998}
\maketitle
\begin{abstract}
{\bf Abstract.}
The harmonic measure (or diffusion field) near a critical percolation
cluster in two dimensions (2D) is considered. Its moments, summed over the
accessible external hull, exhibit a multifractal (Mf) spectrum, which I
calculate exactly. The generalized dimensions $D\left( n\right) $ as well as the
Mf function $f\left( \alpha \right) $ are derived from generalized
conformal invariance, and are shown to be identical to those of the harmonic
measure on 2D random or self-avoiding walks. An exact application to the
impedance of a rough percolative electrode is given. The numerical checks
are excellent. Another set of multifractal exponents is obtained exactly for $n$ independent self-avoiding walks anchored at the random fractal boundary of a percolation cluster.
\end{abstract}
Percolation theory, whose tenuous fractal structures, called incipient clusters, 
present fascinating properties, has served as an archetypal model for critical 
phenomena\cite{stauffer}. The subject has recently enjoyed renewed interest: the 
scaling (continuum) limit
has fundamental properties, e.g., conformal invariance, which present a 
mathematical challenge\cite{langlands,ai1,ben}. Almost uncharted territory in 
exact fractal studies is the {\it harmonic measure}, i.e., the diffusion or 
electrostatic field near an equipotential random fractal boundary, whose 
self-similarity 
is reflected in a {\it multifractal} (Mf) behavior of the harmonic measure 
\cite{BBE}.
  
Mf exponents for the harmonic measure of fractals are especially
important in two contexts: diffusion-limited aggregation (DLA) and the double
layer impedance at a surface. In DLA, the harmonic
measure actually determines the growth process and its scaling
properties are intimately related to those of the
cluster itself\cite{halsey5}. The double layer impedance at a rough surface 
between a good conductor
and an ionic medium presents an anomalous frequency dependence, which has been 
observed by
electrochemists for decades. It was recently proposed that this is at heart
a multifractal phenomenon, directly linked with the harmonic measure of the 
rough 
electrode\cite{halsey7}.  
In both the above contexts, percolation clusters have been studied numerically 
as 
generic models. 
 
In this Letter, I consider incipient percolation clusters in two dimensions 
(2D), 
and determine analytically the exact multifractal exponents of their harmonic 
measure. I use recent advances in conformal
invariance (linked to quantum gravity), which allow for the mathematical 
description
of random walks interacting with other random fractal structures, such as
random walks \cite{duplantier4,lawler}, and self-avoiding walks
\cite{duplantier5}. A further difficulty here is the presence of a subtle geometrical structure in 
the percolation cluster hull, recently elucidated by Aizenman et al.\cite{DAA}.   Excellent agreement with decade-old numerical data is 
obtained, thereby confirming the relevance of conformal invariance to 
multifractality; the exact prediction for the anomalous exponent of a 
percolative 
electrode given here also corroborates the multifractal nature of the latter. As an illustration of the flexibility of the method, I also give the set of exact multifractal exponents corresponding to the average $n$th moment of the probability for a {\it self-avoiding} walk to escape from a percolation cluster boundary.

Consider a two-dimensional very large incipient cluster ${\cal C}$, at the 
percolation threshold $p_{c}$. Define 
$H\left( w\right) $ as the probability that a random walker (RW) launched from 
infinity, {\it first} hits the outer (accessible) percolation hull ${\cal 
H}({\cal C})$ at point $w \in {\cal H}({\cal C})$. We are especially 
interested in the moments of $H$, averaged over all
realizations of RW's and ${\cal C}$ 
\begin{equation}
{\cal Z}_{n}=\left\langle \sum\limits_{w\in {\cal H}}H^{n}\left( w\right)
\right\rangle ,  \label{Z}
\end{equation}
 where $n$ can be, {\it a priori},
a real number. For very large clusters ${\cal C}$ and hulls ${\cal H}\left( 
{\cal 
C}\right) 
$ of average size $R,$ one expects these moments to scale as 
\begin{equation}
{\cal Z}_{n}\approx \left( a/R\right) ^{\tau \left( n\right) },  \label{Z2}
\end{equation}
where $a$ is a microscopic cut-off, and where the multifractal scaling exponents 
$\tau \left(
n\right) $ encode generalized dimensions $D\left( n\right)$, $\tau \left( 
n\right) 
=\left( n-1\right) D\left( n\right) ,$
which vary in a non-linear way with $n$\cite{bb,hent,frisch,halsey1}. Several 
{\it 
a priori} results are known. $D(0)$ is the Hausdorff dimension of the support of 
the
measure. By construction, $H$ is a normalized probability measure, so
that $\tau (1)=0.$ Makarov's theorem \cite{mak}, here applied to the
H\"{o}lder regular curve describing the hull \cite{ai2}, gives the {\it non 
trivial} information dimension $\tau ^{\prime }\left( 1\right) =D\left( 1\right) 
=1.$
The multifractal formalism \cite{bb,hent,frisch,halsey1} further involves
characterizing subsets ${\cal H}_{\alpha }$ of sites of the hull ${\cal H}$
by a Lipschitz--H\"{o}lder exponent $\alpha ,$ such that their local
H-measure scales as 
$H\left( w\in {\cal H}_{\alpha }\right) \approx \left( a/R\right) ^{\alpha }.$
The ``fractal dimension'' $f\left( \alpha \right) $ of the
set ${\cal H}_{\alpha }$ is given by the symmetric Legendre transform of $%
\tau \left( n\right) :$%
\begin{equation}
\alpha =\frac{d\tau }{dn}\left( n\right) ,\quad \tau \left( n\right)
+f\left( \alpha \right) =\alpha n,\quad n=\frac{df}{d\alpha }\left( \alpha
\right) .  \label{alpha}
\end{equation}

Because of the ensemble average (\ref{Z}), values of $%
f\left( \alpha \right) $ can become negative for some domains of $\alpha $ 
\cite{cates}.

This Letter is organized as follows: I first present in detail the findings and 
their potential physical significance and applications, before proceeding with 
the 
more abstract mathematical derivation.

My results for the generalized harmonic dimensions for percolation are 
\begin{equation}
D\left( n\right) =\frac{1}{2}+\frac{5}{\sqrt{24n+1}+5},\quad n\in \left[ -%
{\textstyle{1 \over 24}}%
,+\infty \right) ,  \label{dn}
\end{equation}
valid for all values of moment order $n,n\geqslant -\frac{1}{24}.$
The Legendre transform (\ref{alpha}) of $\tau \left( n\right) =\left( n-1\right) 
D(n)$ reads  
\begin{equation}
\alpha =\frac{d\tau }{dn}\left( n\right) =\frac{1}{2}+\frac{5}{2}\frac{1}{%
\sqrt{24n+1}},  \label{alpha1}
\end{equation}
and 
\begin{equation}
f\left( \alpha \right) =\frac{25}{48}\left( 3-\frac{1}{2\alpha -1}\right) -%
\frac{\alpha }{24},\quad \alpha \in \left( 
{\textstyle{1 \over 2}}%
,+\infty \right) . \label{f}
\end{equation}
\begin{figure}
\centerline{\epsfig{file=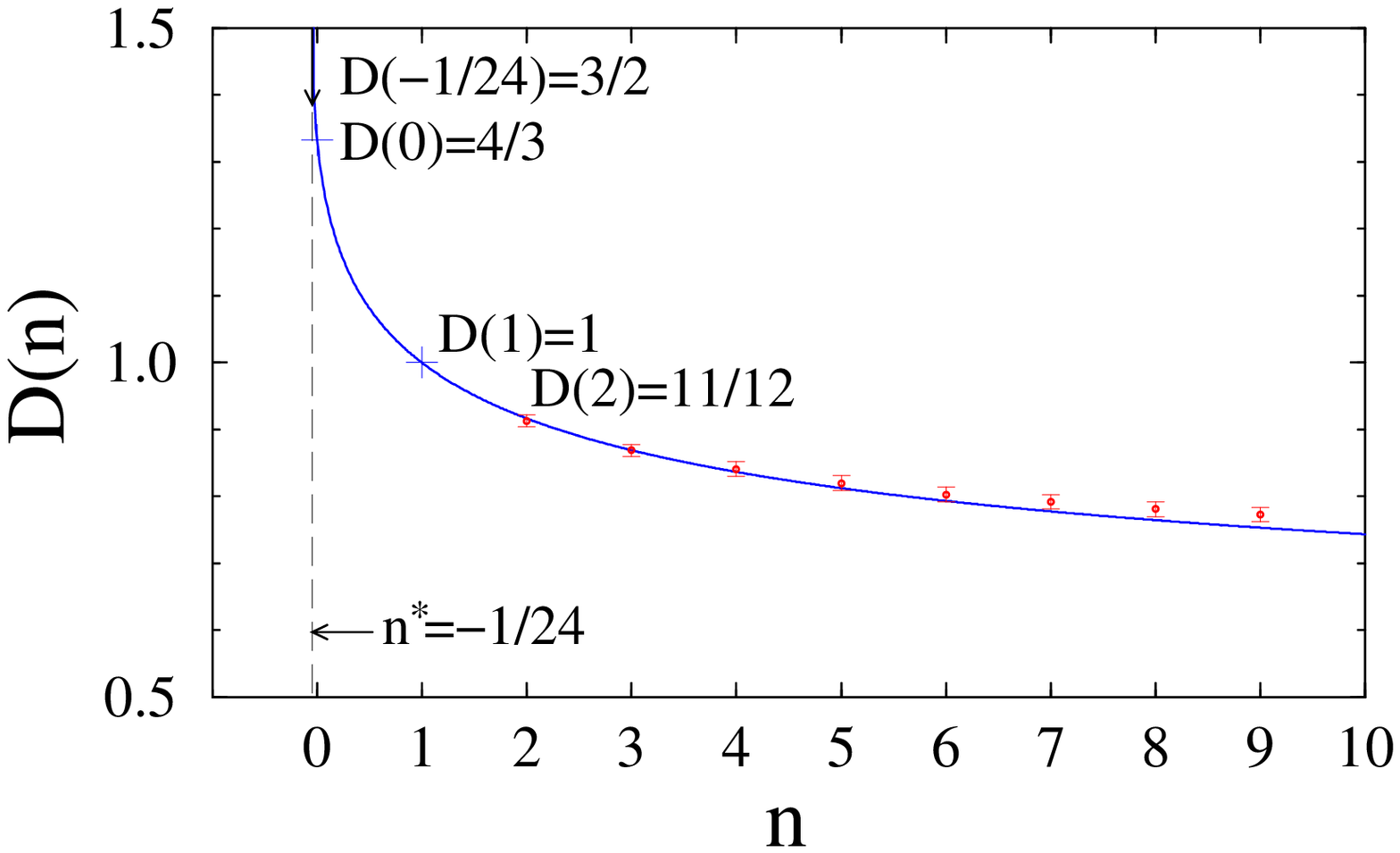,width=8.5cm}}
\end{figure}
\begin{figure}
\vskip -23pt
\centerline{\epsfig{file=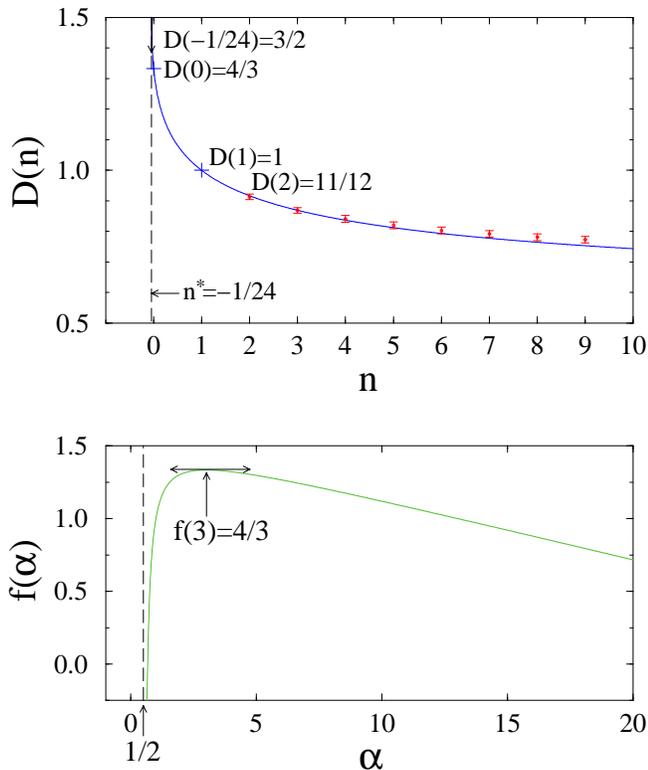,width=8.5cm}}
\smallskip
\caption{Universal harmonic multifractal dimensions $D(n)$, and spectrum
$f(\alpha)$ of a 2D incipient percolation cluster, compared to numerical 
results by Meakin et al. (in red).}
\label{Figure1}
\end{figure}
Figure 1 shows the exact curve $D\left( n\right) $ (\ref{dn}) together with the 
numerical results for $n\in \{2,...,9\} $ by Meakin et al.\cite{meakin}, showing 
fairly good agreement. The slight upwards move from the theoretical curve at 
high 
values of $n$ suggests a
difference between annealed and
apparent quenched averages, as in the DLA case \cite{halsey3}.

The first
striking observation is that the dimension of the support of the measure 
$D\left( 
0\right) \neq D_{{\rm H}},$ where $D_{{\rm %
H}}=\frac{7}{4}$ is the Hausdorff dimension of the standard hull, i.e., the 
outer
boundary of critical percolating clusters \cite{SD}. In fact, $D(0)=\frac{4}{3}$ is the dimension $D_{\rm EP}$ of the {\it accessible external 
perimeter}\cite{GA,DAA}, the other hull sites being located in deep
fjords, which are not probed by the harmonic measure. Its exact value $D_{{\rm %
EP}}=\frac{4}{3}$ has been recently derived in terms of relevant scaling operators describing {\it path crossing} statistics in percolation\cite{DAA}. In the {\it %
scaling continuous} regime of percolation, the fjords {\it do} close, yielding a
{\it smoother} (self-avoiding) accessible
perimeter of dimension $\frac{4}{3}$. 
This is in
agreement with the instability phenomenon observed numerically on a lattice: 
removing the fjords with narrow necks causes a discontinuity of the effective 
dimension of the hull from $D_{{\rm H}}\simeq \frac{7}{4}$ to $D_{%
{\rm EP}}\simeq \frac{4}{3},$ whatever microscopic restriction rules are 
choosen\cite{GA}. In other respects, a 2D polymer at the $\Theta$-{\it point} is 
known 
to obey exactly the statistics of a percolation hull\cite{DS}, and the Mf 
results 
(\ref{dn}-\ref{f}) therefore apply {\it also} to that case.  

An even more striking fact is the complete identity of Eqs. (\ref{dn}-\ref{f}) 
to 
the corresponding results {\it both} for random
walks and self-avoiding walks (SAW's) \cite{duplantier5}. In particular, $%
D\left( 0\right) =\frac{4}{3}$ is the Hausdorff dimension of a SAW, common to  
the 
{\it external frontier} of a percolation hull and of a Brownian 
motion\cite{duplantier4,lawler}. Seen from outside, these three fractal curves
are not distinguished by the harmonic measure. As we shall see, this fact is 
linked to the
presence of a universal underlying conformal field theory with a vanishing 
central
charge $c=0.$

The singularity at $\alpha=\frac{1}{2}$ in the multifractal function $f(\alpha)$ is due to points on the fractal boundary where the latter has the local geometry of a needle. Indeed, by elementary conformal covariance, a local wedge of opening angle $\theta$ yields an electrostatic potential, i.e., harmonic measure, which scales as $H(R) \sim R^{-\frac{\pi}{\theta}}\sim R^{-\alpha}$, thus, formally, $\theta=\frac{\pi}{\alpha}$, and $\theta=2\pi$ corresponds to the lowest possible value $\alpha=\frac{1}{2}$.
The linear asymptote of the $%
f\left( \alpha \right) $ curve for $\alpha \rightarrow +\infty ,f\left(
\alpha \right) \sim -\frac{\alpha }{24}$ corresponds to the
lowest part $n\rightarrow {n^{\ast}} =-\frac{1}{24}$ of the spectrum of 
dimensions. Its linear shape is quite reminiscent of the case of a 2D DLA 
cluster 
\cite
{ball}. Define ${\cal N}\left( H\right)$ as the  number of sites having a 
probability $H$ to be hit. Using the Mf formalism to change from variable $H$ to  
$%
\alpha $ (at fixed value of $a/R)$, shows that ${\cal N}\left( H\right)$ obeys, 
for
$H\rightarrow 0,$ a power law behavior  
with an exponent 
$\tau ^{\ast }=1+%
\mathrel{\mathop{\lim }\limits_{\alpha \rightarrow +\infty }}%
\frac{1}{\alpha }f\left( \alpha \right)=1+n^{\ast}.$  
Thus we predict
\begin{equation}
{\cal N}\left( H\right)|_{H\rightarrow 0}\approx H^{-{\tau}^{\ast}},   \tau 
^{\ast}=\frac{23}{24}. 
\label{nh}
\end{equation}
This $\tau ^{\ast }=0.95833...$ compares very well with the result 
$\tau ^{\ast }=0.951\pm 0.030$, obtained for $%
10^{-5}\leqslant H\leqslant 10^{-4}$ \cite{meakin}.

Let us consider for a moment the different, but related, problem of the {\it 
double layer impedance}
of a {\it rough} electrode. In some range of frequencies $\omega $, the
impedance contains an anomalous ``constant
phase angle'' (CPA) term $\left( i\omega
\right) ^{-\beta }$, where $\beta <1$. It was believed that $\beta $ would be 
solely determined by the
Hausdorff dimension ${D\left( 0\right)}$ of the electrode surface. From a 
natural 
RW representation of the impedance, a different scaling law was recently 
proposed: 
$\beta =\frac{D\left( 2\right) }{D\left( 0\right) }$  
\label{beta}
(here in 2D), where $D\left( 2\right) $ is the multifractal dimension of the 
H-measure on the rough electrode\cite{halsey7}. In the case of a
2D porous percolative electrode, our
results (\ref{dn}) give $D\left( 2\right)
\equiv \frac{11}{12},$ $D\left( 0\right)=\frac{4%
}{3}$, whence $\beta =\frac{11}{16}=0.6875.$ This
compares very well with a numerical RW algorithm result\cite{MS}, which yields  
an 
effective CPA exponent $\beta \simeq 0.69,$
nicely vindicating the multifractal description\cite{halsey7}.

Let me
 now give the main lines of the derivation of exponents $D\left(
n\right) $ by generalized {\it conformal invariance}. We focus on site 
percolation 
on the 2D triangular lattice; by universality the results are expected to apply 
to 
other $2D$ (e.g., bond) percolation models. The boundary lines of the
percolation clusters, i.e., of connected sets of occupied hexagons, form
self-avoiding lines on the
dual hexagonal lattice. 
They obey the statistics of loops in the ${\cal O}\left( N=1\right)$ model, 
where 
$N$ is the loop fugacity, in the so-called ``low-temperature phase'', a fact we 
shall recover below\cite{SD}. 

By the very definition of the H-measure, $n$ independent RW's diffusing away 
from 
the hull give a geometric representation of the $n^{th}$
moment $H^{n},$ for $n$ {\it integer}. The values so derived for $n\in {\Bbb %
N}$ will be enough, by convexity arguments, to obtain the analytic
continuation for arbitrary $n$'s. Figure 2 depicts $n$ independent random walks, 
in a bunch, {\it %
first} hitting the external hull of a percolation cluster at a site $w=\left(
\bullet \right) .$ As explained in ref.\cite{DAA}, such a site, to belong to the {\it accessible} hull, must 
remain, in the 
{\it continuous scaling limit},  the
source of at least {\it three non-intersecting crossing paths}, noted ${\cal 
S}_{3},$ 
reaching to a
(large) distance $R.$ These paths are ``{\it monochromatic}'': 
one path runs only through occupied (light blue) sites; the other two,
{\it dual\/} lines, 
run through  empty (white) sites\cite{DAA}. 
The definition of the {\it standard} hull requires only the 
origination, in the scaling limit, of a ``bichromatic'' pair of lines ${\cal 
S}_2$\cite{DAA}.
Points lacking additional dual lines are not accessible to RW's
after the scaling limit is taken, because
their (white) exit path becomes a strait pinched by other parts of the
(light blue) occupied cluster. The bunch of independent RW's avoids
the occupied cluster, and defines its own envelope as a set of two
{\it boundary }lines separating it from the occupied part of the
lattice, thus from ${\cal S}_{3}$ (Fig. 2).

Let us introduce the notation $A\wedge B$ for two sets, $A$, $B$, of random 
paths, conditioned to
be {\it mutually avoiding, }and{\it \ }$A\vee B$ for two {\it independent}, thus 
possibly
intersecting, sets \cite{duplantier5}. Now consider $n$
independent RW's, or Brownian paths ${\cal B}$ in the scaling limit, in a bunch 
noted $\left(
\vee {\cal B}\right) ^{n},$ {\it avoiding} a set ${\cal S}_{\ell }\equiv \left( 
\wedge 
{\cal P}\right) ^{\ell }$ of $\ell $ {\it non-intersecting} crossing paths in 
the
percolation system. Each of the latter paths passes only through occupied sites, or only through empty ({\it dual}) ones. The probability that the Brownian and percolation paths altogether traverse the 
annulus ${\cal %
D}\left( a, R\right) $ from the inner boundary circle of radius $a$ to the outer 
one at distance $R$, i.e., are in a ``star'' configuration ${\cal S}_{\ell 
}\wedge 
\left( \vee {\cal B}\right) ^{n}$(Fig. 2), is expected to scale for $%
R/a\rightarrow \infty $ as 
\begin{equation}
{\cal P}_{R}\left( {\cal S}_{\ell }\wedge n\right) \approx
\left( a/R\right) ^{x\left( {\cal S}_{\ell }\wedge n\right) },  \label{xp}
\end{equation}
where we used ${\cal S}_{\ell }\wedge n \equiv {\cal S}_{\ell }\wedge \left( 
\vee 
{\cal B}\right) ^{n}$ as a short hand notation, and where $x\left( {\cal 
S}_{\ell 
}\wedge n\right) $ is a new critical exponent
depending on $\ell $ and $n.$ It is convenient to introduce similar {\it %
``surface}'' probabilities $\tilde{{\cal P}}_{R}\left( {\cal S}_{\ell }\wedge
n\right) \approx \left( a/R\right) ^{\tilde{x}\left( {\cal S}_{\ell }\wedge
n\right) }$ for the same star configuration of paths, now crossing through the 
half-annulus $\tilde{{\cal D}}\left( a, R\right) $ in the {\it half-plane}.
\begin{figure}
\centerline{\epsfig{file=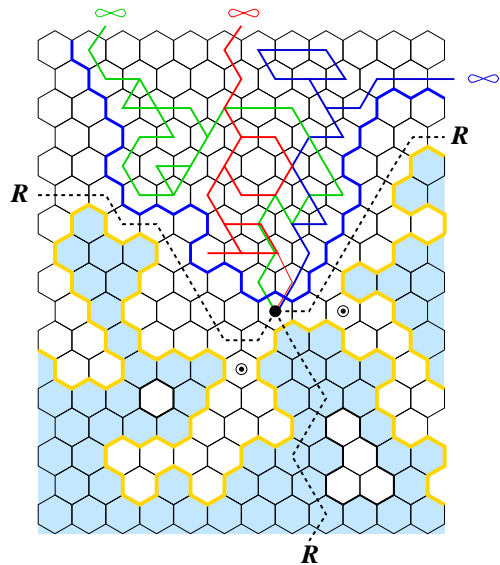,width=6.5cm}}
\smallskip
\caption{An ``active'' site $(\bullet )$ on the accessible external
perimeter for site percolation on the triangular lattice. It is
defined by the existence, in the {\it scaling limit}, of $\ell =3$
non-intersecting, and ``monochromatic'' crossing paths ${\cal S}_{3}$ (dotted
lines), one on the incipient (light blue) cluster, the other two on
the dual empty (white) sites. The points $\odot$ are entrances of
fjords, which close in the scaling limit and won't support the
harmonic measure. Point $(\bullet )$ is first reached by three
independent RW's (red, green, blue), contributing to $H^3 (\bullet
)$. The hull of the incipient cluster (golden line) avoids the outer
frontier of the RW's (thick blue line). A (random) Riemann map of the
latter onto the real line ${\Bbb R}$ 
reveals the presence of an underlying $\ell=3$ path-crossing {\it boundary} 
operator, i.e, of an $L=4$-line,
or two-cluster, boundary operator in the associated ${\cal O}(N=1)$
model, with dimension in the half-plane
$\tilde{x}_{\ell=3 }=\tilde{x}_{L=\ell +1=4}^{{\cal O}\left( N=1\right) }=\tilde
{x}^{{\cal C}}_{k=2}=2.$ Both accessible hull and Brownian
paths have a frontier dimension $\frac{4}{3}$.}
\label{Figure2}
\end{figure} 
When $n \to 0, {\cal P}_{R}\left( {\cal S}_{\ell }\right) \left( \text{resp. 
}\tilde{{\cal P}}%
_{R}\left( {\cal S}_{\ell }\right) \right) $ is the probability
of having $\ell $ simultaneous monochromatic(non-intersecting) path-crossings traversing the 
annulus in the plane (resp. half-plane), with associated exponents $x_{\ell }\equiv x\left( {\cal S}_{\ell } \wedge 
0\right) $ and $\tilde{x}_{\ell }\equiv \tilde{x}\left( {\cal S}_{\ell } \wedge 
0\right)$\cite{ai1,DAA}. These exponents have been studied in ref.\cite{DAA}, and shown rigorously to be actually independent of the coloring of the paths, with the restriction in the bulk that there exist at least a path on occupied sites and one on dual ones, thus $\ell>1$. Here the exponents appear as analytic continuations of the harmonic measure ones (\ref{xp}) to $n \to 0$, and should correspond to the definitions of ref.\cite{DAA}.
  
In terms of definition (\ref{xp}), the harmonic measure moments (\ref{Z}) simply scale as
${\cal Z}_{n}\approx R^2{\cal P}_{R}\left( {\cal S}_{\ell =3}\wedge n\right)$
\cite{cates}, 
 which, combined with Eqs. (\ref{Z2}) and (\ref{xp}), leads to 
\begin{equation}
\tau \left( n\right) =x\left( {\cal S}_{3}\wedge n\right) -2.  \label{tt}
\end{equation}

Using the fundamental mapping of the conformal field theory (CFT) in the {\it 
plane} $%
{\Bbb R}^{2},$ describing a critical statistical geometrical system, to the
CFT on a fluctuating abstract random Riemann surface, i.e., in presence of {\it 
quantum gravity} \cite{KPZ}, I have recently shown that there exist two
universal functions $U,$ and $V,$ depending only on the central charge $c$ of 
the CFT, which suffice to generate all geometrical exponents involving
{\it mutual avoidance} of random {\it star-shaped} sets of paths of the critical 
system
 \cite{duplantier5}. For 
$c=0,$ which corresponds to RW's, SAW's, and {\it percolation}, these
universal functions are:
\begin{eqnarray}
U\left( x\right) &=&\frac{1}{3}x\left( 1+2x\right) , \hskip2mm V\left( x\right) 
=\frac{1}{24}\left( 4x^{2}-1\right) .  \label{U}
\end{eqnarray}
with $V\left( x\right) \equiv U\left( 
{\textstyle{1 \over 2}}%
\left( x-%
{\textstyle{1 \over 2}}%
\right) \right)$. Consider now two arbitrary\ random sets $A,B,$ involving each 
a 
collection of paths in a star configuration, with proper scaling crossing
exponents $x\left( A\right) ,x\left( B\right) ,$ or, in the half-plane, crossing 
exponents $\tilde{x}\left( A\right) ,\tilde{x}\left(
B\right) .$ If one fuses the star centers and requires $A$ and $B$ to stay
mutually avoiding, then the new crossing exponents, $x\left( A\wedge
B\right) $ and $\tilde{x}\left( A\wedge B\right) ,$ obey the {\it star
algebra} \cite{duplantier4,duplantier5} 
\begin{eqnarray}
x\left( A\wedge B\right) &=&2V\left[ U^{-1}\left( \tilde{x}\left( A\right)
\right) +U^{-1}\left( \tilde{x}\left( B\right) \right) \right]  \nonumber \\
\tilde{x}\left( A\wedge B\right) &=&U\left[ U^{-1}\left( \tilde{x}\left(
A\right) \right) +U^{-1}\left( \tilde{x}\left( B\right) \right) \right] ,
\label{x}
\end{eqnarray}
where $U^{-1}\left( x\right) $ is the inverse function of $U$
\begin{equation}
U^{-1}\left( x\right) =\frac{1}{4}\left( \sqrt{24x+1}-1\right) .  \label{u1}
\end{equation}

If, on the contrary, $A$ and $B$ are {\it independent} and can overlap, then
by trivial factorization of probabilities, $x\left( A\vee B\right)
=x\left( A\right) +x\left( B\right) ,$ and $\tilde{x}\left( A\vee B\right) =%
\tilde{x}\left( A\right) +\tilde{x}\left( B\right)$ \cite{duplantier5}. The 
rules 
(\ref{x}), which mix bulk and boundary exponents, can be understood as simple 
factorization properties on a random Riemann surface, i.e., in quantum gravity 
\cite{duplantier4,duplantier5}, or as 
recurrence relations in ${\Bbb R}^{2}$ between conformal Riemann maps of the 
successive mutually avoiding paths onto the line ${\Bbb R}$\cite{lawler}. 
On a random surface, $U^{-1}\left( \tilde{x} \right)$ is the boundary
dimension corresponding to the value $\tilde{x}$ in ${\Bbb R} \times
{\Bbb R}^{+}$, and the sum of $U^{-1}$ functions in Eq. (\ref{x})
represents linearly the juxtaposition $A \wedge B$ of two sets of
random paths near their random frontier, i.e., the product of two
``boundary operators'' on the random surface. The latter sum is mapped
by the functions $U$, $V$, into the scaling dimensions in ${\Bbb
R}^2$ \cite{duplantier5}. The structure thus unveiled is so stringent
that it immediately yields the values of the percolation crossing exponents
$x_{\ell}, \tilde {x}_{\ell}$ of ref.\cite{DAA}, and our harmonic measure exponents
$x\left( {\cal S}_{\ell }\wedge n\right) $ (\ref{xp}).  First, for a
set ${\cal S}_{\ell }=\left( \wedge {\cal P}\right) ^{\ell }$ of $\ell
$ crossing paths, we have from the recurrent use of (\ref{x})
\begin{equation}
x_{\ell }=2V\left[ \ell U^{-1}\left( \tilde{x}_{1}\right) \right] ,\quad 
\tilde{x}_{\ell }=U\left[ \ell U^{-1}\left( \tilde{x}_{1}\right) \right] .
\label{xl}
\end{equation}
For percolation, two values of half-plane crossing exponents $\tilde{x}_{\ell }$ 
are known by
{\it elementary} means: $\tilde{x}_{2}=1,\tilde{x}_{3}=2\cite{ai1,DAA}.$ From (%
\ref{xl}) we thus find $U^{-1}\left( \tilde{x}_{1}\right) 
=\frac{1}{2}U^{-1}\left( 
\tilde{x}_{2}\right) =\frac{1}{3}U^{-1}\left( \tilde{x}_{3}\right) =\frac{1}{%
2},$ (thus $ \tilde{x}_{1}=\frac{1}{3}$ \cite{ca}), which in turn gives 
\[
x_{\ell }=2V\left( 
{\textstyle{1\over 2}}%
\ell\right) =\frac{1}{12}\left(
{\ell}^{2}-1\right), 
\tilde{x}_{\ell }=U\left(
{\textstyle{1\over 2}}%
\ell\right)=\frac{\ell }{6}\left( \ell +1\right).
\]
We thus recover the identity, previously rigorously established in ref.\cite{DAA}, of $x_{\ell }=x_{L=\ell }^{{\cal O}\left( 
N=1\right) }, \tilde{x}_{\ell }=\tilde{x}_{L=\ell +1}^{{\cal O}\left( N=1\right) 
}$ with the $L$-line exponents of the associated ${\cal O}\left( N=1\right)$ loop 
model, in the ``low-temperature phase''. 
For $L$ {\it even}, these exponents also govern the existence of 
$k=\frac{1}{2}L$ 
{\it spanning} clusters \cite{SD,DAA}, with the identity $x_{k}^{\cal C}=x_{\ell =2k}=%
\frac{1}{12}\left( 4k^{2}-1\right) $ in the bulk\cite{SD}, and $\tilde{x}_{k}^{\cal C}=%
\tilde{x}_{\ell =2k-1}=\frac{1}{3}k\left( 2k-1\right) $ in the half-plane 
\cite{SD,D7,ai3}. 
The non-intersection exponents
of $k'$ {\it Brownian paths} are also given by $x_{\ell}$, $\tilde{x}_{\ell}$ for 
${\ell}=2k'$\cite{duplantier4}, so we observe a {\it complete} equivalence 
between 
a Brownian path
and {\it two} percolating crossing paths, in both the plane and half-plane.

For the harmonic exponents in (\ref{xp}), we fuse the two objects 
${\cal 
S}_{\ell}$ and $\left( \vee {\cal B}\right) ^{n}$ into a new star ${\cal 
S}_{\ell}\wedge n $ (see Fig. 2), and use (\ref{x}). We just have seen that the 
boundary $\ell$-crossing exponent of ${\cal S}_{\ell}$, $\tilde{x}_{\ell}$,  
obeys 
$U^{-1}\left( \tilde{x}_{\ell}\right) =\frac{1}{2}\ell.$ The bunch of $n$ 
independent Brownian paths have their own  half-plane crossing exponent 
$\tilde{x}\left( \left( \vee {\cal B}\right) ^{n}\right) =n\tilde{x}%
\left( {\cal B}\right) =n,$ since the boundary dimension of a single Brownian 
path
is trivially $\tilde{x}%
\left( {\cal B}\right)=1$\cite{duplantier4}. Thus we obtain
\begin{equation}
x\left( {\cal S}_{\ell}\wedge n\right) =2V\left(
{\textstyle{1\over 2}}%
\ell 
+U^{-1}\left( n\right) \right).  \label{fina}
\end{equation}
Specifying to the case $\ell=3$ finally gives from (\ref{U})(\ref{u1}) 
\[
x\left( {\cal S}_{3}\wedge n\right) =2+\frac{1}{2}\left( n-1\right) +\frac{5%
}{24}\left( \sqrt{24n+1}-5\right) ,
\]
from which $\tau \left( n\right) $ (\ref{tt}), and $D\left( n\right) $ Eq.(%
\ref{dn}) follow, {\bf QED}.

This formalism immediately allows many generalizations. For instance, in place of $n$ random walks, one can consider a set of $n$ {\it independent self-avoiding} walks $\cal P$, which avoid the cluster fractal boundary, except for their common anchoring point. The associated multifractal exponents $ x\left( {\cal S}_{\ell}\wedge ( \left( \vee {\cal P}\right) ^{n}\right)$ are given by the same formula (\ref{fina}), with the argument $n$ in $U^{-1}$ simply replaced by the boundary scaling dimension of the bunch of independent SAW's, namely \cite{duplantier5} $\tilde{x}\left( \left( \vee {\cal P}\right) ^{n}\right) =n\tilde{x}%
\left( {\cal P}\right) =n\frac{5}{8}$, for $\ell=3.$ 
These exponents govern the universal multifractal behavior of the $n$th moments of the probability that a self-avoiding walk escapes from the random fractal boundary of a percolation cluster in two-dimensions. I thus find that they are identical to those obtained when the random fractal boundary is taken as the frontier of a Brownian path or a self-avoiding walk \cite{duplantier5}.

{\bf Acknowledgements}: I thank M. Aizenman, D. Kosower, and T. C. Halsey for fruitful discussions.

\end{document}